\documentstyle{mn}
\input{epsf}
\begin{document}
 

\def\ax{\alpha_{\rm X}}
\def\g{$\gamma$}
\def\zm{z_{\rm max}}
\def\nh{N_{\rm H}}
\def\nlo{\langle n L\rangle_0}
\def\af{A_{\rm Fe}}
\def\taut{\tau_{\rm T}}
\def\ec{E_{\rm c}}
\def\ginga{{\it Ginga}}
\def\asca{{\it ASCA}}
\def\exosat{{\it EXOSAT}}
\def\iue{{\it IUE}}
\def\luv{L_{\rm UV}}
\def\fuv{F_{\rm UV}}
\def\fbb{F_{\rm bb}}
\def\fabs{F_{\rm abs}}
\def\le{L_{\rm E}}
\def\lxg{L_{{\rm X}\gamma}}
\def\fxg{F_{{\rm X}\gamma}}
\def\granat{{\it GRANAT}}
\def\as{\alpha_{\rm soft}}
\def\rosat{{\it ROSAT}}
\def\ee{e$^\pm$}
\def\rs{r_{\rm S}} 
\def\rc{r_{\rm c}} 
\def\rxg{r_{{\rm X}\gamma}}

\hyphenation{Max-well-ian brems-strahl-ung syn-chro-tron
black-body ap-pen-dix i-so-tro-pic com-pact-ness com-pact-nes-ses}
 
\def\ast{\mathchar"2203} \mathcode`*="002A
\newbox\grsign \setbox\grsign=\hbox{$>$} \newdimen\grdimen \grdimen=\ht\grsign
\newbox\simlessbox \newbox\simgreatbox \newbox\simpropbox
\setbox\simgreatbox=\hbox{\raise.5ex\hbox{$>$}\llap
     {\lower.5ex\hbox{$\sim$}}}\ht1=\grdimen\dp1=0pt
\setbox\simlessbox=\hbox{\raise.5ex\hbox{$<$}\llap
     {\lower.5ex\hbox{$\sim$}}}\ht2=\grdimen\dp2=0pt
\setbox\simpropbox=\hbox{\raise.5ex\hbox{$\propto$}\llap
     {\lower.5ex\hbox{$\sim$}}}\ht2=\grdimen\dp2=0pt
\def\simgreat{\mathrel{\copy\simgreatbox}}
\def\simless{\mathrel{\copy\simlessbox}}
\def\simprop{\mathrel{\copy\simpropbox}}

\topmargin=-1cm
 
\title[The UV/X-ray correlation in NGC 4151]
{The origin of the correlation between the UV and X-rays in NGC 4151} 

\author[A. A. Zdziarski and P. Magdziarz]
{\parbox[]{6.in} {Andrzej A. Zdziarski$^{1,2}$ and Pawe\l\ Magdziarz$^3$} \\
$^1$N. Copernicus Astronomical Center, Bartycka 18, 00-716 Warsaw, Poland,
Internet: aaz@camk.edu.pl\\
$^2$Institute for Theoretical Physics, University of California, Santa Barbara,
CA 93106, USA\\
$^3$Astronomical Observatory, Jagiellonian University, Orla 171, 30-244 
Cracow, Poland, Internet: pavel@camk.edu.pl\\} 
 
\date{Accepted 1996 January 9. Received 1995 December 28; 
in original form 1995 September 14}

\maketitle
\begin{abstract}
We propose that the UV/X-ray correlation observed in NGC 4151 by \exosat\/ and 
\iue\/ is due to reemission of the X-ray flux absorbed by cold clouds in the 
line of sight. This transmission model satisfies the energy balance, provides 
a good fit to the X-ray and UV data, and predicts no Compton reflection of 
X-rays (as observed). 
The correlation can alternatively be explained by reprocessing of X-rays and 
\g-rays by a cold accretion disk. A specific model in which X-rays and \g-rays 
are emitted by a dissipative patchy corona and have the spectrum cut off at 
$\sim 100$ keV (as seen by OSSE) is shown to be energetically possible even if 
it does not give as satisfactory a fit as the transmission model. However, 
the model predicts a Compton-reflection spectral component, whose presence 
remains to be tested by future detectors sensitive above 10 keV. 
\end{abstract} 

 
\section{INTRODUCTION AND SUMMARY}
\label{s:intro}

Correlation between the UV and X-ray fluxes in radio-quiet AGNs ranges from 
good to non-existent (e.g., Ulrich 1994). So far, a good correlation was 
observed in the Seyfert galaxies NGC 4151 (Perola et al.\ 1986, hereafter P86) 
and NGC 5548 (Clavel et al.\ 1992). Here we consider the origin of the 
correlation in NGC 4151, a nearby Seyfert 1.5 galaxy at the distance of 
$D\simeq 20$ Mpc. 

The correlation in NGC 4151 was observed during two \iue/\exosat\/ campaigns 
in 1983 November 7--19 and 1984 December 16--1985 January 2 (P86). The UV flux 
at 8.5 eV was proportional to the 2--10 keV X-ray flux corrected for 
absorption (Section \ref{s:data}). The time delay between the flux variations 
was less than $\sim 2$ days. The delay, however, was more tightly constrained 
to $\simless 0.5$ day in 1993 December (Warwick et al.\ 1996), when the light 
curves were sampled at time intervals much shorter than in 1983--86.  

In the first part of this {\it Letter}, we present a new model explaining the 
correlation. We propose that the UV radiation is from reemission of X-rays 
absorbed by cold matter outside the X-ray source (Section \ref{s:abs}). Thus, 
we identify the matter responsible for the UV emission with that responsible 
for the observed absorption of X-rays. The column density, $\nh \sim 10^{23}$ 
cm$^{-2}$, is determined from modeling the X-ray spectra (P86). We show that 
this model (hereafter referred to as the transmission model) gives an 
excellent fit to the UV and X-ray data as well as it satisfies the energy 
balance. The model implies only residual Compton reflection (Lightman \& White 
1988) because the absorber is Thomson-thin, $\tau\sim 0.1$, which is in 
agreement with hard X-ray observations (Maisack \& Yaqoob 1991; Yaqoob et al.\ 
1993; Zdziarski, Johnson \& Magdziarz 1996, hereafter Z96). The absorbed flux 
is mostly from the observed 2--10 keV range and thus the model does not depend 
on the shape of the \g-ray spectrum, which was not observed during the 
X-ray/UV observations. 

From considering the absorber column density, ionization balance, and 
free-free absorption we determine that the absorber is not uniform but rather 
made of many dense discrete clouds. The clouds are optically thick to UV 
radiation and their temperature is $kT\simeq 3$ eV. The  obtained cloud 
parameters are similar to those postulated to exist in the inner regions of 
AGNs by Guilbert \& Rees (1988) and Ferland \& Rees (1988). 

After presenting the transmission model we reexamine a previously proposed 
model of reprocessing by an optically-thick accretion disk (hereafter referred 
to as the reflection model; Section \ref{s:repro}). In that model, the 
correlation is due to the disk irradiated from above by X-rays and \g-rays 
(hereafter X-\g) and reemitting the absorbed flux in the UV (Ulrich et al.\ 
1991; Ulrich 1994; Perola \& Piro 1994, hereafter PP94). 

A number of problems appear in the reflection model. One is the observed 
absence of a strong spectral component at $\simgreat 10$ keV expected from 
Compton reflection of the irradiating X-rays from the optically thick disk 
(Yaqoob et al.\ 1993; Z96). [This contrasts the other Seyfert with a UV/X-ray 
correlation, NGC 5548, which X-ray spectrum does have a prominent 
Compton-reflection component (Nandra et al.\ 1991).] Furthermore, the most 
detailed version of the model, proposed by PP94, requires that in order to 
satisfy the energy balance the extrapolated X-ray emission has to extend to 
$\sim 4$ MeV, which has not been confirmed by recent OSSE measurements (Z96). 
In that case, the observed large intrinsic dispersion of the 2--10 keV spectral 
index (P86) implies that the extrapolated power-law flux near $\sim 4$ MeV 
(which dominates the absorbed flux for the hard spectrum of NGC 4151) becomes 
a strongly dispersed function of the 2--10 keV flux, which in turn implies a 
poor X-ray/UV correlation (see Section \ref{s:repro}), contrary to the 
data. 

We show here that most of the above difficulties can be dealt with by choosing 
a source geometry different from the central optically-thick spherical 
emission region proposed by PP94. Namely, we propose that the X-\g\ source 
forms a dissipative corona above the surface of the disk, as in the model of 
Seyfert 1's of Haardt \& Maraschi (1993) and Haardt, Maraschi \& Ghisellini 
(1994). However, the corona is more patchy than in Seyfert 1's to account for 
X-ray spectra of NGC 4151 being much harder than those of average Seyfert 1's 
(Z86). We find that the corona geometry results in a strong reduction of the 
required X-\g\ luminosity with respect to the model of PP94. We find that the 
UV/X-\g\ energy balance is satisfied even if the incident X-\g\ spectrum is 
cut off at $\sim 100$ keV, as observed by OSSE. This low cutoff energy also 
reduces the dispersion in the correlation of the total absorbed power with the 
2--10 keV flux. However, the model does predict a Compton-reflection spectral 
component only marginally consistent with hard X-ray observations. 

Note that both models are similar in that both postulate absorption of X-rays 
by some cold medium, which subsequently reprocesses and reemits the absorbed 
flux as UV photons. The principal difference is that the medium is 
Thomson-thin in the transmission model whereas it is very Thomson-thick in the 
reflection model. 

\section{The data}
\label{s:data}

We have refitted the \exosat\/ ME argon data (from 1.2 to 10--15 keV) using a 
model consisting of a power law, Fe K$\alpha$ line, absorption by a dual 
neutral absorber (i.e., a product of partial and complete covering, P86), and 
a constant soft X-ray component modeled as in Z96. We assume an Fe 
overabundance of 2.6, which is the average value obtained by P86. We find the 
X-ray energy spectral indices, $\ax$ ($F_E\propto E^{-\ax}$, where $F_E$ is 
the energy flux), are weakly constrained by the argon data alone. Those 
indices were determined before in the 3.5--25 keV range by P86, who used the 
ME xenon data extending up to 25 keV in addition to the argon data. The 
addition of the xenon data resulted in the allowed ranges of $\ax$ much 
narrower than ours. Since the xenon ME data are no longer available we used 
the ranges of $\ax$ as given in P86 (in the range of $\sim 0.2$--0.7), which 
we found consistent with the argon data. (We have checked that relaxing this 
assumption increases the error bars shown on the figures here but it does not 
affect our qualitative conclusions.) 

\begin{figure}
\centering
\epsfxsize=8.4cm \epsfbox{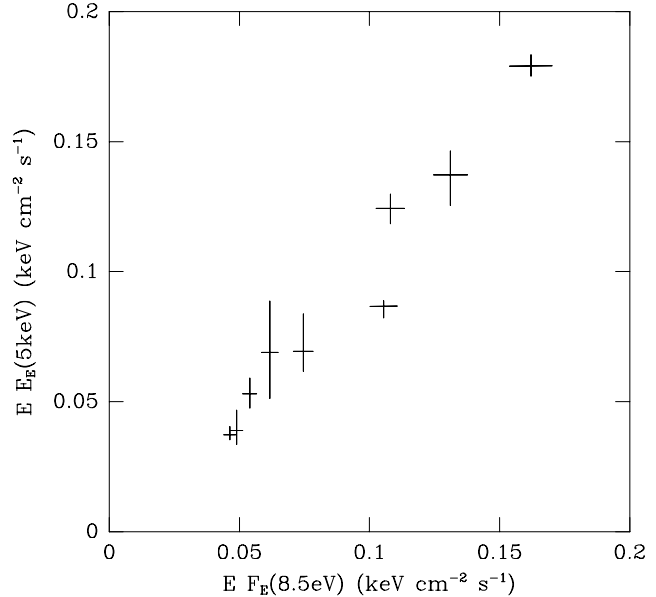}
\caption{The correlation between the absorption-corrected $EF_E$ at 5 keV and 
at 8.5 eV observed during the 1983--1986 campaigns by \exosat\/ and \iue. The 
vertical error bars on Figs.\ 1, 2, 3a are for $\Delta\chi^2=2.7$, i.e., 90 
per cent confidence for one parameter of interest. 
 }
\label{fig:exiue}
\end{figure}

Fig.\ \ref{fig:exiue} shows the resulting correlation between the 
absorption-corrected $EF_E$ at 5 keV (of the power-law only) and at 8.5 eV 
[from \iue, dereddened with $E(B-V)=0.05$, PP94; Kriss et al.\ (1995) obtain a 
similar value of 0.04]. We see the correlation is close to linear at high 
significance. The values of $EF_E$ in X-rays are similar to those in the UV, 
which is indeed highly suggestive for the origin of the UV from some form of 
reprocessing of X-rays. 

\section{The transmission model}
\label{s:abs}

\begin{figure}
\centering
\epsfxsize=7cm \epsfbox{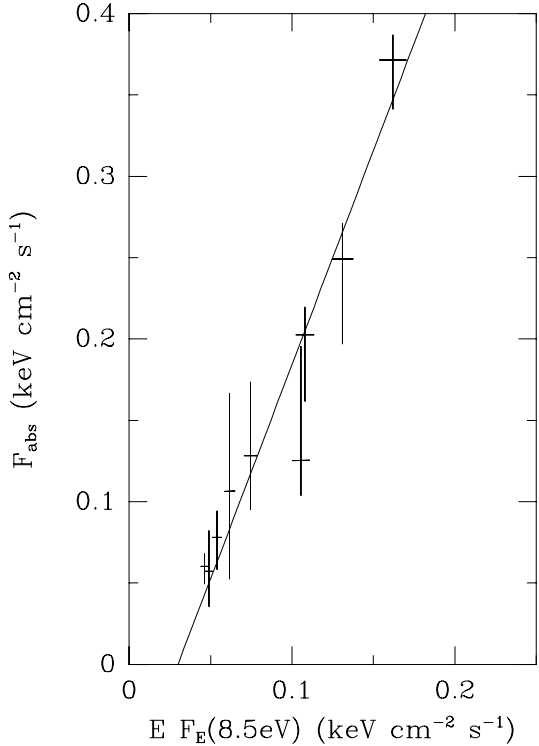}
\caption{Comparison of the prediction of the transmission model with the 
\iue/\exosat\/ data. The crosses give the absorbed flux, $\fabs$, obtained 
from the \exosat\/ data, and $EF_E(8.5$ eV) is from \iue. The solid line gives 
$EF_E(8.5$ eV) as a function of $\fabs$ from the transmission model (see 
text). 
 }
\label{fig:abs}
\end{figure}

The fits in Section \ref{s:data} give us the fluxes absorbed by the cold 
matter in the line of sight, $\fabs$. We assume that the absorbing clouds are 
dense enough for almost complete thermalization of that flux (see below). We 
determine the temperature of the absorber by requiring that the spectral index 
in the UV between 8.5 and 7.2 eV equals the average observed value, $\alpha=-
0.15$ (Ulrich et al.\ 1991; PP94). This gives $kT\simeq 3$ eV. Fig.\ 
\ref{fig:abs} compares the predictions of that model with the data, assuming 
 \begin{equation}
\label{eq:F}
\fuv=f \fabs+F_0, 
\end{equation}
 where $\fuv$ is the integrated UV flux (related to the plotted flux at 8.5 eV 
by the blackbody formula), $F_0=0.05$ keV cm$^{-2}$ s$^{-1}$ is a residual 
flux, and $f=0.60$ takes into account a covering of the X-\g\ source by the 
clouds of $<4\pi$ as well as an efficiency ($<1$) of the conversion of the 
absorbed flux into the blackbody continuum. We see that there is a very good 
agreement between the data and the model in spite of our rather simplified 
assumptions  (constant $T$, $f$, $F_0$, pure blackbody UV spectrum). Since 
$f<1$, the model does satisfy the energy balance, i.e., there is more than 
enough power in the absorbed X-rays to account for the observed UV emission. 

We consider now constraints on the absorbing medium. For simplicity, we assume 
that all the absorption occurs in a single cloud. From the fits, we obtain the 
typical column density of a cloud in the partial absorber of $\nh \sim 
10^{23}$ cm$^{-2}$ with the typical covering factor of $\sim 0.5$. The $\nh$ 
of the complete absorber is a few times less. The dominant thermalization 
process at 8.5 eV is free-free, which implies that the size of a single cloud, 
$\rc$, is $\simless 10^7$ cm (equivalently $n\simgreat 10^{16}$ cm$^{-3}$), in 
order for the free-free optical depth at energies $\simless 10$ eV to be $>1$ 
at $T$ and $\nh$ as above. At those conditions, collisions alone are 
sufficient to almost fully ionize hydrogen, and thus there are enough free 
electrons for free-free absorption. 

The average size of the entire absorber, $r$, can be determined from the 
blackbody law, $\luv$ ($\sim 10^{43}$ erg s$^{-1}$) $=4\pi \sigma T^4 r^2$, to 
be $\sim 10^{14}$ cm, which satisfies the limit $r\simless 10^{15}$ cm (for 
geometry close to spherical) from the relative UV/X-ray time delay (Section 
\ref{s:intro}). The ionization parameter, $L/(nr^2)$, is then $\simless 0.1$, 
which implies metals are almost neutral, and thus soft X-rays are strongly 
absorbed in bound-free processes. The parameters of the absorber are similar 
to those studied by Ferland \& Rees (1988; see their Fig.\ 8) and Guilbert \& 
Rees (1988), for which they find the UV spectrum at $\simless 10$ eV is indeed 
thermalized. Equipartition magnetic fields can confine the clouds with the 
above parameters (Celotti, Fabian \& Rees 1992). The X-\g\ source has to cover 
$\sim 0.1$ of the sky as seen from a typical cloud in order to satisfy the 
X-ray constraints on soft-photon starvation (Z96). Summarizing, the 
transmission model both explains the data and appears to be self-consistent. 
 
It appears that accretion of the clouds in the observed absorber alone is 
insufficient to power the AGN. The total mass in the clouds is independent of 
their number, and is $\sim f(4\pi/3) \nh r^2 m_{\rm p}\sim 10^{28}$ g. On the 
other hand, the mass required to account for $L\sim (3$--$10)\times 10^{43}$ 
erg s$^{-1}$ (assuming a cutoff at $\sim 100$ keV as observed by OSSE) at an 
efficiency of 0.1 is $\sim (3$--$10)\times 10^{27}/\beta_{\rm r}$ g, where 
$\beta_{\rm r}$ is the radial velocity. Thus, the two masses marginally agree 
only for highly relativistic radial velocities. 

\section{The reflection model} 
\label{s:repro}

Here we reexamine the reflection model, in which the correlation arises from 
X-\g\ irradiating an optically-thick accretion disk (e.g., PP94). The incident 
X-{\g} flux is mostly absorbed since the albedo is $A\sim 0.3$ (Z96). The 
absorbed power is reprocessed and reemitted locally by the disk mostly as UV 
blackbody radiation, similarly as in the transmission model. 

PP94 have developed a detailed model explaining the correlation, in which the 
X-\g\ source forms an optically thick central sphere irradiating a surrounding 
disk. They find that the amplitude of the UV emission at 8.5 eV and the 
7.2--8.5 eV spectral index (Ulrich et al.\ 1991) can be both explained only 
when the hard power-law emission extends to rather hard \g-rays; e.g., a 
cutoff above $\sim 4$ MeV is required for a typical $\alpha\sim 0.5$. (We have 
reproduced those results using the assumptions of PP94.) 

As discussed in Section \ref{s:intro}, there are major difficulties with such 
hard \g-ray emission. First, it has not been seen by the modern \g-ray 
detectors, OSSE and SIGMA (e.g., Z96; Finoguenov et al.\ 1995). Although 
\g-rays were not measured during the \exosat/\iue\/ campaign, a UV/X-ray 
correlation was also seen in 1993 December (see Section \ref{s:other}; Warwick 
et al.\ 1996), when the OSSE data showed $kT\simeq 50^{+20}_{-10}$ keV (Z96). 
Second, $\ax$ varies from 0.2 to 0.7 in the \exosat\/ data (P86). As a result 
of the dispersion of $\ax$, the total fluxes, $\fxg$, in the power-law spectra 
extrapolated to 4 MeV correlate very poorly with the 2--10 keV fluxes. We find 
that indeed the corresponding reprocessed fluxes are away by factors up to 3 
from their best-fit linear correlation with the UV, which invalidates the 
model (unless the dispersion is compensated for by suitable shapes of the 
\g-ray spectra, which seems unlikely). 

However, it is possible that other source geometries could be more compatible 
with the data. We study here a model in which the X-\g\ source forms a patchy 
corona above the surface of an accretion disk (Haardt, Maraschi \& Ghisellini 
1994). We assume that all dissipation takes place in the corona (Haardt \& 
Maraschi 1993; Svensson \& Zdziarski 1994). Thus, the local power, $Q(r)$, 
released in the hot corona depends on radius, $r$, as the disk dissipation 
rate (Shakura \& Sunyaev 1973), 
 \begin{equation}
Q(r) = {3\over 8\pi} {G M \dot M\over r^3} \left[ 1-\left( 3\rs\over 
r\right)^{1/2} \right],
\label{eq:Q}
\end{equation}
 where $\rs =2GM/c^2$ is the Schwarzschild radius, $M$ is the black hole mass, 
and $\dot M$ is the mass accretion rate. The power released in the corona is 
emitted partly outward and partly towards the disk. We assume here that the 
luminosity intercepted by the disk equals $R$ times the luminosity emitted 
outward (and observed). $R$ can be $<1$ due to, e.g., a finite size of the 
cold disk or an outflow of the hot plasma. On the other hand, the luminosity 
emitted outward is related to the total observed X-\g\ flux, $\fxg$, by $\lxg= 
4\pi D^2\fxg$. Then the luminosity absorbed and reemitted thermally by the 
disk is, 
 \begin{equation}
\luv= (1-A) R\lxg=4\pi (1-A) R D^2\fxg,
\label{eq:L}
\end{equation}
 where $A$ ($\simeq 0.3$) is the albedo. Energy balance then implies that the 
luminosity $\luv$ equals the disk emission at the rate, 
 \begin{equation}
\sigma [T(r)]^4=Q(r) (1-A) R/(1+R),
\label{eq:T}
\end{equation}
 integrated over the disk surface. This allows us to solve for the constant, 
$M\dot M$, in equation (\ref{eq:Q}), as well as it shows that the UV spectrum 
is a standard blackbody disk spectrum (Shakura \& Sunyaev 1973) (rather than 
the single-temperature blackbody spectrum in the transmission model, or the 
spectrum from irradiation by an optically thick sphere in the model of PP94). 

\begin{figure}
\centering
\epsfxsize=8.4cm \epsfbox{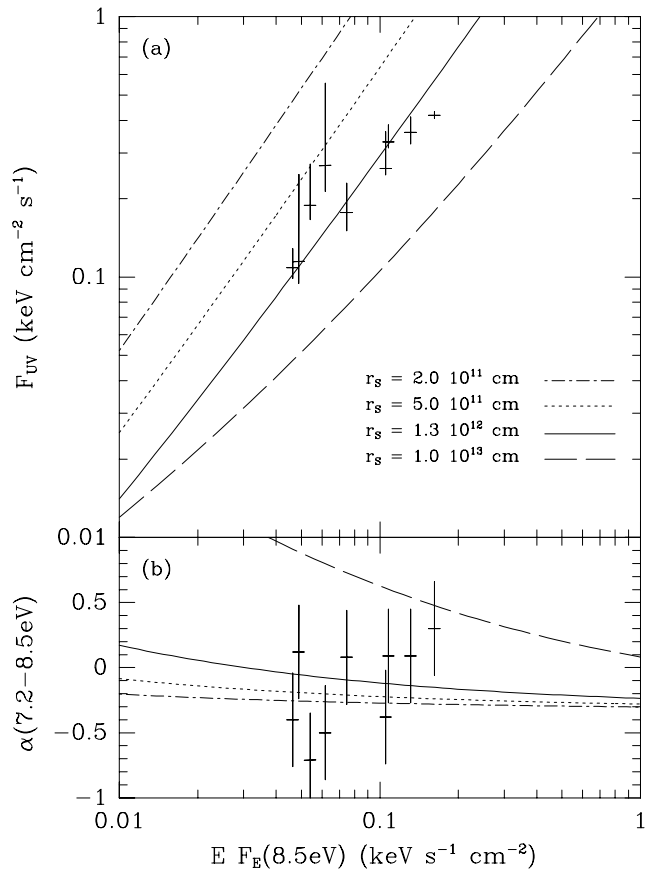}
\caption{Comparison of the predictions of our reflection model with the data. 
{\it (a)} Crosses give the total UV flux, $\fuv$, from reemission of the 
absorbed X-\g\ flux as obtained extrapolating the \exosat\/ X-ray power laws 
to \g-rays assuming thermal Comptonization at $kT=60$ keV, see text. The 
curves relate the total $\fuv$ to $E F_E(8.5\,{\rm eV})$, as predicted by the 
disk spectrum for various Schwarzchild radii, $\rs$. {\it (b)} Crosses give 
the UV spectral indices. Curves give the indices predicted by the disk-corona 
model. We see that the {\it (solid)} curve for $\rs=1.3\times 10^{12}$ cm 
provides the best description of both the UV fluxes and spectra. 
 }
\label{fig:repro}
\end{figure}

The total observed UV flux is given by, 
 \begin{equation}
\fuv= 2\mu(1-A)R F_{{\rm X}\gamma},
\label{eq:fuv}
\end{equation}
 where the factor $2\mu$ accounts for disk emission with a constant specific 
intensity.  Using the blackbody disk spectrum, we can then relate the observed 
8.5 eV flux to the total UV flux, $\fuv$, as a function of $\rs^2 \mu$, where 
$\mu=\cos i$ and $i$ is the disk viewing angle (see PP94). Fig.\ 
\ref{fig:repro}{\it a\/} shows this relation for several values of $\rs$ 
assuming $\mu=0.42$ [as obtained by {\it HST}, Evans et al.\ (1993); $\rs$ for 
other values of $\mu$ scales as $\mu^{-1/2}$]. Fig.\ \ref{fig:repro}{\it b\/} 
shows the corresponding 7.2--8.5 eV spectral indices, compared with the 
observed values (Ulrich et al.\ 1991). 

The curves in Fig.\ \ref{fig:repro}{\it a\/} can be compared with the total 
fluxes, $\fuv$, absorbed by the disk and reemitted in the UV as obtained 
extrapolating the \exosat\/ power laws to \g-rays in the way observed by OSSE, 
i.e., for thermal Comptonization at $kT =60$ keV (Z96). Z96 obtained the best 
fit values of $R\sim 0$--0.3 (at $\mu=0.42$) based on measuring a Compton 
reflection component in hard X-ray spectra, with $R=0.5$ marginally consistent 
with the data (see also Maisack \& Yaqoob 1991), and we thus use $R=0.5$. Note 
that $R\ll 1$ would not be consistent with the assumption that reprocessing is 
in the disk. Crosses in Fig.\ \ref{fig:repro}{\it a\/} give $\fuv$ for 
$R\mu=0.21$ (corresponding to $R=0.5$ and $\mu=0.42$). 

We can now compare predictions of the disk model for various $\rs$ with the UV 
spectral indices and $\fuv$ inferred from the \exosat\/ data. We find first 
that $\rs<2.4 \times 10^{12}$ cm fit the UV spectral indices, with the upper 
limit corresponding to $\Delta\chi^2=2.7$. The value of $\rs=1.3\times 
10^{12}$ cm provides the best fit to the absorbed fluxes inferred from the 
X-ray data for $R\mu=0.21$. The upper limit on $\rs$ corresponds to 
$R\mu=0.15$. We note that the relative contribution of Compton reflection to 
the hard X-ray spectra decreases {\it slower\/} than $\propto \mu$ (Magdziarz 
\& Zdziarski 1995), and thus the problem of the weakness of Compton reflection 
together with the presence of reprocessing in NGC 4151 cannot be resolved by 
postulating $\mu\ll 1$ (Z96). 

Note that the fit in Fig.\ \ref{fig:repro}{\it a\/} is significantly worse 
than that for the transmission model (Fig.\ \ref{fig:abs}). This is caused by 
the main contribution to $\fuv$ being from $\sim 100$ keV, relatively far away 
from the 2--10 keV range, where the linear correlation is observed. Still, the  
correlation is much closer to linear than that obtained by extrapolating the 
X-ray spectra to 4 MeV. 

The range of $\rs$ obtained above corresponds to $M\simless 8\times 10^6 
M_\odot$. At the largest $M$ and the source size $\simgreat 10\rs$, $\Delta 
t\simgreat 10^3$ s, compatible with the observed longer variability time 
scales. (Note that $M$ larger than that given above can be obtained for a Kerr 
metric.) The range of the maximum disk temperatures is 7--10 eV. Note that the 
implied range of $EF_E$ at 100 keV is 0.1--0.3 keV cm$^{-2}$ s$^{-1}$, which 
is much larger than that in 1991--93, when OSSE observed a relatively constant 
100 keV flux (Z96). 

Summarizing this section, we find that reprocessing by a disk is a viable 
model for NGC 4151 if the X-\g\ spectra are cut off at $\sim 100$ keV. Our 
conclusion differs from that of PP94, who required that the power-law spectra 
extend to $\sim 4$ MeV in order to satisfy the energy balance. The difference 
is entirely explained by the different geometrical models adopted by PP94 and 
us. First, our geometry gives $2(1+\mu)$ times more reprocessed flux  (for 
unit observed X-\g\ flux). Physically, this is due to a small solid angle 
subtended by the disk as seen from a point on the surface of the sphere, as 
well as due to the optically-thick sphere emitting only outside (as opposed to 
the corona emitting both outside and inside). Second, our model gives the UV 
spectrum different from that of PP94, with our UV spectrum compatible with the 
data without the need for \g-ray emission harder than observed. 

The main {\it caveat\/} for this model is the predicted Compton reflection, 
with $R\mu \simgreat 0.15$. Such reflection is only marginally allowed by the 
X-ray data.

\section{OTHER OBSERVATIONS}
\label{s:other} 

The two models proposed above explain the UV/X-ray correlation observed in 
1983--86. However, the behavior of the UV and X-\g\ in NGC 4151 changes from 
one epoch to another. In 1979 May 19--31, $EF_E (8.5$ eV) was $\sim 0.25$ keV 
cm$^{-2}$ s$^{-1}$ while $EF_E(5\,{\rm keV})$ was $\sim 0.07$--0.12 keV cm$^{-
2}$ s$^{-1}$ (Perola et al.\ 1982). The UV flux is larger than that implied by 
the specific correlation of Fig.\ \ref{fig:exiue}. The sampling of the UV flux 
was too sparse to  either establish  or rule out a UV/X-ray correlation in 
that period with an $F_0$ larger than that in 1983--86. 

A correlation with a large $F_0$ has been in fact observed in 1993 November 
30--December 10 (Warwick et al.\ 1996). The $EF_E(8.5\,{\rm eV})$ varied 
within $\sim 0.5$--0.6 keV cm$^{-2}$ s$^{-1}$ [using $E(B-V)=0.05$, Section 
\ref{s:data}] together with the 1--2 keV flux (measured by \rosat\/ and 
\asca). The $EF_E(5\,{\rm keV})$ varied in the $\sim 0.05$--0.25 keV cm$^{-2}$ 
s$^{-1}$ range, whereas the flux above 50 keV measured simultaneously by OSSE 
varied only within $\pm 15$ per cent. 

Preliminary estimates indicate that 1993 December data can be explained by 
either the transmission or reflection model; discussion is given in Warwick et 
al.\ (1996). We only note that a large residual UV flux is required in the 
transmission model whereas the reflection model may explain most of the 
constant UV component as due to disk reprocessing of the approximately 
constant \g-rays. 

\section{CONCLUSIONS}
\label{s:dis} 

We have found that the X-ray/UV correlation in NGC 4151 seen by \exosat\/ and 
\iue\/ can be explained by either of two models. In the first one, the UV flux 
is due to reemission of the X-ray flux absorbed by small, Thomson-thin clouds, 
which absorption is directly measured in X-rays. The model provides an 
excellent fit to the data. 

In the other model, the X-\g\ source forms a patchy, dissipative corona above 
the surface of an accretion disk. The disk reprocesses a fraction of the X-\g\ 
flux into the UV, which has been proposed before but for a central X-\g\ 
source. The geometry proposed here allows the UV/X-\g\ energy balance to be 
satisfied for X-\g\ spectra cut off at $\sim 100$ keV, as observed by OSSE. 

The reflection model requires spectral hardening above 10 keV,  
owing to the presence of Compton reflection. 
This can be tested by X-\g\ observations including coverage of the 
10--50 keV range (e.g., by {\it XTE}), where the relative contribution from 
Compton reflection peaks. The two models make different predictions regarding 
the specific UV-X-\g\ variability, and thus can be tested by coordinated 
multiwavelength observations. 

Note that the transmission model produces the UV within the X-ray absorber so 
that the expected absorption in the UV is different than that produced in 
the reflection model, where the UV and X-rays both are affected by the same 
absorber (as proposed by Warwick, Smith \& Done 1995). The recent data 
from the Hopkins Ultraviolet Telescope appear to rule out the latter 
possibility (Kriss et al.\ 1995), thus favouring the transmission model. 

\section*{ACKNOWLEDGMENTS}
This research has been supported in part by the Polish KBN grant 2P03D01008, 
NASA {\it GRO} and {\it XTE} grants, and the NSF grant PHY94-07194. 
It has made use of data obtained 
through the High Energy Astrophysics Science Archive Research Center Online 
Service, provided by NASA/GSFC. We are grateful to T. Kallman, G. Madejski, 
J. Miko\l ajewska, L. Piro, R. Svensson, R. Warwick and P. \.Zycki for 
valuable discussions, and to C. Done for detailed comments on this work.

\end{document}